\documentclass[a4paper,aps,twoside,brazil,english,prd,amsmath,nofootinbib,
superscriptaddress,showpacs
]{revtex4-1}

\usepackage[T1]{fontenc}
\usepackage[utf8]{inputenc}
\usepackage[brazil,english]{babel}
\usepackage[usenames,rgb,dvipsnames]{xcolor}
 \definecolor{darkblue}{RGB}{0,0,150}

\usepackage{array}
\usepackage{booktabs}
\usepackage{amssymb}
\usepackage{tensor}
\usepackage{graphicx}
\usepackage{subfig}
\usepackage{nicefrac}
\usepackage{amsthm}
\usepackage{mathrsfs}

\usepackage[unicode]{hyperref}
\usepackage{cancel}
\usepackage{setspace}
\usepackage{ulem}
\hypersetup{
  pdfinfo = {
    Author = {Alan, Morgan, Mimoso},
    Title = {MTS junction}
  },
  colorlinks = true,
  breaklinks = true,
 linkcolor = blue,
 urlcolor = blue,
 citecolor = blue,
}
\usepackage{multirow}
\usepackage{enumerate}
\usepackage{listings}
\usepackage[
            linecolor=orange,
            backgroundcolor=red!30,
            colorinlistoftodos]{todonotes}
\usepackage{comment}

\newcommand{\ud}{\ensuremath{\mathrm{d}}}
\newcommand{\Lie}{\ensuremath{\mathcal{L}}}

\DeclareMathAlphabet{\mathpzc}{T1}{pzc}{m}{it}

\def\coxa{{\Huge
C\kern-.1667em\lower.5ex\hbox{O}\-X\kern-.1667em\lower.5ex\hbox{A}\@}%
\index{CoXa}
}


\begin{document}
\title{Matching collapse and expansion across Matter Trapping surfaces in inhomogeneous $\Lambda$CDM models}
\author{Alan Maciel}
\email{alan.silva@ufabc.edu.br}
\thanks{ORCID: \href{https://orcid.org/0000-0002-1919-2140}{0000-0002-1919-2140}}
\address{Centro de Matem\'atica, Computa\c c\~ao e Cogni\c c\~ao, Universidade Federal do ABC,\\
 Avenida dos Estados 5001, CEP 09210-580, Santo Andr\'e, S\~ao Paulo, Brazil}
\author{M. Le Delliou}
\email{delliou@lzu.edu.cn, Morgan.LeDelliou.IFT@gmail.com}
 \thanks{ORCID: \href{https://orcid.org/0000-0003-3655-2547}{0000-0003-3655-2547}}
\affiliation{-Institute of Theoretical Physics \& Research Center of Gravitation, Lanzhou University, Lanzhou 730000, China\\
	-Key Laboratory of Quantum Theory and Applications of MoE, Lanzhou University, Lanzhou 730000, China\\
	-Lanzhou Center for Theoretical Physics \& Key Laboratory of Theoretical Physics of Gansu Province, Lanzhou University, Lanzhou 730000, China} 	
	\affiliation{Instituto de Astrof\'isica e Ci\^encias do Espa\c co, Universidade de Lisboa,
	Faculdade de Ci\^encias, Ed.~C8, Campo Grande, 1769-016 Lisboa, Portugal}
 \affiliation{Universit\'e de Paris-Cit\'e, APC-Astroparticule et Cosmologie (UMR-CNRS 7164), 
 F-75006 Paris, France.}
 \author{José P. Mimoso}
\email{jpmimoso@fc.ul.pt,jpmimoso@ciencias.ulisboa.pt}
\thanks{ORCID: \href{https://orcid.org/0000-0002-9758-3366}{0000-0002-9758-3366}}
\address{Departamento de F\'{i}sica and Instituto de  Astrof\'{i}sica e Ci\^encias do Espa\c co, \\ 
Faculdade de Ci\^{e}ncias da Universidade de Lisboa, \\ 
Campo Grande, Ed. C8 1749-016 Lisboa,
Portugal}

\begin{abstract}
 In previous works, Matter Trapping Surfaces (MTS) were defined as hypersurfaces separating cosmologically expanding regions of spacetime from regions where collapse can take place independently.
In the present work we examine the MTS, for the restriction to spherical dust plus $\Lambda$, proving that it actually is a characteristic surface of the Cauchy problem (generated by its characteristic curves), which opens the possibility for infinite solutions. This translate as the MTS being a boundary between arbitrarily independent solutions, reminiscent of the Birkhoff theorem effects.
This property is illustrated with combinations of 3 examples containing MTSs and $\Lambda$ ($\Lambda$CDM, Schwarzschild-de\,Sitter, Lema\^itre-Tolman-Bondi-de\,Sitter: LTBdS -- i.e. the inhomogeneous, spherically symmetric $\Lambda$CDM). The LTBdS model presents a static, stable MTS for the first time.
\end{abstract}
\maketitle
\today
\section{Introduction}

The initial data problem formulation of General Relativity (GR) is the key to interpret what seems a purely geometric theory into a well defined physical theory that dictates the evolution of an admissible system. Since in GR what constitutes an initial instant is not uniquely defined, we deal with a Cauchy problem, where the conditions are imposed on spacelike hypersurfaces.

In previous works \cite{MLeDM09,LeDMM09a,Mimoso:2013iga}, we defined the concept of Matter Trapping Surfaces (MTS), which are dynamically defined in spherically symmetric spacetimes as surfaces that separate the regions of expansion and contraction of the spacetime. The MTS indicates the emergence of a bound region where the dynamics is decoupled from cosmological expansion, determining the frontier between the dominance of global and local physics.

This distinction is relevant for many astrophysical and cosmological fundamental questions. For instance: in which regimes can Newtonian physics provide an adequate description, and when must one rely on the full machinery of General Relativity? How do overdense regions dynamically decouple from the overall expansion of the universe? The concept of MTS offers a physically motivated criterion to distinguish these opposed regimes, by objectively characterizing the transition between locally bound and globally expanding domains.

Yet an important question that needs formal  clarification is whether the MTS can be used as a boundary  that would entirely determine the dynamics
on each side in well posed form. This would be the case if the Einstein's Field Equations (EFEs) together with the MTS can be recast as a Cauchy problem.

As widely appreciated the initial value data and Cauchy problem are questions that are crucial for a well posed problem, and have attracted a great deal of interest in the field.  The fundamental, and seminal  works on this issue can be traced back to the  works of Y. Choquet-Bruhat \cite{Foures-Bruhat:1952grw,Choquet-Bruhat:1969ywq}, A. Lichnérowicz \cite{Lichnerowicz:1944zz,Lichnerowicz1944}, S. Deser \cite{Deser:1967zzb},  and 
J. W. York \cite{York:1971hw,York:1972sj,York:1973ia,York:2004gb}. Other works  have made relevant contributions in connection to particular aspects of this endeavour \cite{Komar:1958ymq,Bartnik:2002cw,Chrusciel:2004cc,Anderson:2000mt,Ringstrom:2015jza}, and a few  enlightening reviews provide a thorough overview of the numerous contributions, namely in what regards the main conceptual questions \cite{Isenberg:2013iva}, and in what concerns the important numerical applications \cite{ Tichy:2016vmv,Gourgoulhon:2007ue}.

In the present work, we impose boundary conditions on the MTS, assuming a matter content composed of dust and a cosmological constant. The result is that, in this case, the MTS corresponds to a characteristic surface in the Cauchy sense, implying that the evolution on each side of this surface is completely decoupled. This means that, apart from physical considerations regarding the continuity of certain quantities such as mass-energy, any interior solution can be matched with any exterior solution across an MTS. This characterizes the MTS not only as a separating surface between regions of expansion and collapse, but also as a shield that protects each side from the detailed dynamics of the other. The only physical variables that each side perceives from the other are the averaged quantities that must be matched at the MTS to ensure continuity.

An outline of the work is as follows,. In Sec.~\ref{sec:LCDM}, we introduce the formal tools  that we shall use in the subsequent analysis of the Cauchy problems, namely the 1+1+2 spacetime splitting, its translation into the GPG line element vocabulary and the MTS characterisation.
We focus on the $\Lambda$-CDM model (i.e. $\Lambda$ plus dust model) that is now favoured by observations and is somewhat the present standard model of cosmology. In this very same section we  develop our analysis of the MTS as a Cauchy problem. In Sec.~\ref{sec:Illustrate} we illustrate our results considering some particular realizations of scenarios where the MTS separates two different solutions. We envisage Schwarzschild-de Sitter matching, the Einstein-de Sitter model, and we build a more evolved model where there is a sigmoid-like transition between two Lemaître-Tolman-Bondi solutions motivated by Large-Scale-Structure (LSS)  models. Finally in Sec.~\ref{sec:conclusion} we conclude and provide final remarks discussing our results. 

\section{The MTS as a Cauchy problem for dust with $\Lambda$}

\label{sec:LCDM} 
We recall the extended Lemaître-Tolman-Bondi (LTB) model, presenting
a spherically symmetric content of dust 
and a cosmological constant, approached as a perfect fluid combination. 
The dust with $\Lambda$ case 
corresponds to the $\Lambda$CDM model presently favoured by cosmology,
and hence presents an interest on its own \cite{2019arXiv191203687L}.
In this setting we consider the MTS as a limit surface, and 
we study the possibility of integrating spacetime from boundary
data given on the MTS towards each side. That is, investigating 
each side of the MTS 
to ascertain whether it can be identified as a Cauchy surface for
the PDE system given by Einstein field equations (EFE) in this setup.


\subsection{From 1+1+2 to GPG}

The 1+1+2 formalism applied to spherically symmetric metrics consists
in choosing a timelike future directed unit vector $n^{a}$ along
the matter flow, and a unit spacelike vector $e^{a}$ orthogonal
to $n^{a}$ and the spheres of symmetry. The metric can be written
as: 
\begin{gather}
g_{ab}=-n_{a}n_{b}+e_{a}e_{b}+N_{ab}\,,\label{eq:metric112}
\end{gather}
where $N_{ab}$ correspond to the induced 2-metric on the spheres
of symmetry. By introducing the areal radius $r$, we can write $N_{ab}=r^{2}\Omega_{ab}$
where $\Omega_{ab}$ is the metric of the unit 2-sphere.

\noindent In order to study the dynamics of those spacetimes we resort
to the use of geometrical scalars defined from those quantities, namely,
the 2-expansions $\Theta_{n}$ and $\Theta_{e}$ defined, when the areal radius can be defined, as 
\begin{gather}
\Theta_{n}=\frac{2}{r}\Lie_{n}r\,,\label{theta_n}\\
\Theta_{e}=\frac{2}{r}\Lie_{e}r\,.\label{theta_e}
\end{gather}

In general we can define $\Theta_{X}$ for any $X^{a}$ in the $(n^{a},e^{a})$ plane by the same formula. The interest here is to relate
this approach 
with 
the well known GPG coordinates. In order to achieve this goal, we
recall the definition of the Misner-Sharp mass-energy $M_{ms}$ \citep[hereafter MS]{MisnerSharp}:
\begin{gather}
g^{ab}\partial_{a}r\partial_{b}r=1-\frac{2M_{ms}}{r}\,,\label{M_ms-def1}
\end{gather}
which, together with Eq.~\eqref{eq:metric112}, gives us 
\begin{gather}
-(n^{a}\partial_{a}r)^{2}+(e^{b}\partial_{b}r)^{2}=1-\frac{2M_{ms}}{r}\,,
\end{gather}
as the areal radius gradient is orthogonal to the 2-spheres. We may rearrange it as 
\begin{gather}
e^{b}\partial_{b}r=\pm\sqrt{1+(n^{a}\partial_{a}r)^{2}-\frac{2M_{ms}}{r}}\,.\label{eq:epartialr}
\end{gather}

We interpret the term $n^{a}\partial_{a}r$ as the fluid radial velocity $\frac{\ud r}{\ud\tau}$.
We are 
then 
motivated to define 
\begin{gather}
E=(n^{a}\partial_{a}r)^{2}-\frac{2M_{ms}}{r}\,,\label{eq:Edefinition}
\end{gather}
which corresponds to twice the Newtonian mechanic energy per unit
mass\footnote{\vspace{-.3cm}\begin{spacing}{0}In some sources $E$ is defined as the Newtonian energy per unit
mass, but here we prefer to avoid a factor 2 appearing in the GPG
metric.\end{spacing}}.

\noindent Using the scalar function $E$, Eq.~\eqref{eq:epartialr}
becomes 
\begin{gather}
e^{b}\partial_{b}r=\pm\sqrt{1+E}\,.\label{eq:epartialrE}
\end{gather}

The GPG coordinates correspond to using a timelike coordinate $t$
chosen along the flow, along with the areal radius as the spacelike
coordinate orthogonal to the spheres of symmetry. Therefore, translating
from the abstract index notation to a coordinate notation we have
\begin{gather}
n_{a}\ud x^a=
-\alpha(t,r)\ud t\,,\label{eq:ntranslation}
\end{gather}
where we chose the $-$ sign in order to guarantee $t$ is future
directed with positive $\alpha(t,r)$, that we recognise as the shift function
in the Arnowitt-Deser-Misner 
decomposition \cite[hereafter ADM]{Arnowitt:1959ah}. 
Since, by construction, $n_{a}e^{a}=0$, $e^{a}$ has the form $(0,e^{r},0,0)$.
By Eq.~\eqref{eq:epartialrE}, we conclude that 
\begin{gather}
e^{r}=\sqrt{1+E}\Rightarrow e^{a}\partial_a=
\sqrt{1+E(t,r)}\partial_{r}\,,
\end{gather}
where we chose the $+$ sign in order to direct $e^{a}$ outwards,
that is, in the direction of increasing $r$.

\noindent In order to translate the full metric into coordinate notation
we have to find the form of $e_{a}$. Since $e^{a}$ is a unit vector,
applying $e^{a}e_{a}=1$ gives us $e_{a}=(e_{t},\frac{1}{\sqrt{1+E}},0,0)$,
with $e_{t}$ unconstrained. Therefore, by defining a function $\beta(t,r)=-e_{t}\sqrt{1+E}$, such that it is positive for outwards flow,
we write: 
\begin{gather}
e_{a}\ud x^a=
\frac{1}{\sqrt{1+E(t,r)}}\left(-\beta(t,r)\ud t+\ud r\right)\,.\label{eq:etranslation}
\end{gather}

Plugging Eqs.~\eqref{eq:ntranslation} and \eqref{eq:etranslation}
into Eq.~\eqref{eq:metric112} and making the final translation 
\begin{gather}
N_{ab}\partial_a\partial_b=
r^{2}(\ud\theta^{2}+\sin^{2}\theta^{2}\ud\varphi^{2})=r^{2}\ud\Omega^{2}\,,
\end{gather}
we obtain the well known GPG line element \cite{Lasky:2006zz,adler-2005,LaskyLun06b,LaskyLun07,Gautreau:1984PhRvD186}
: 
\begin{gather}
\ud s^{2}=-\alpha(t,r)^{2}\ud t^{2}+\frac{(-\beta(t,r)\ud t+\ud r)^{2}}{1+E(t,r)}+r^{2}\ud\Omega^{2}\,.\label{eq:GPGmetric}
\end{gather}

\noindent 

In order to interpret the meaning of $\beta(t,r)$, we compute $n^{a}$
in GPG coordinates 
\begin{gather}
n^{a}\partial_a=
\frac{1}{\alpha(t,r)}\left(\partial_{t}+\beta(t,r)\partial_{r}\right)\,,
\end{gather}
which gives us 
\begin{gather}
n^{a}\partial_{a}r=\frac{\beta(t,r)}{\alpha(t,r)}\,,
\end{gather}
and therefore relates $\beta$ with the fluid areal radial velocity.

Thus the lapse function $\alpha(t,r)$ regulates the evolution between
adjacent hypersurfaces, while the shift vector $\beta(t,r)\,\partial_{r}$
corrects point to point for the flow so that the radial coordinate
remains the areal radius $r$.


\subsection{Dynamical Equations}

We 
consider the perfect fluid case, and formulate the Einstein field
equations in terms of the Misner-Sharp mass $M_{ms}$, and of the
energy function $E$, respectively defined in (\ref{M_ms-def1}) and
(\ref{eq:Edefinition}). This allows us to single out the MTS in a
way that makes clear how it constrains the matter flow, before 
recasting the dynamical equations in Cauchy form for the restriction
to dust and $\Lambda$. 
The solution may be given associating $M_{ms}$ with the cosmological
constant and
perfect fluid masses, and introducing the contribution of the cosmological
constant at the level of the shift solution.


Choosing $n^{a}$ as the fluid rest observer 4-velocity,
\footnote{\vspace{-.3cm}\begin{spacing}{0}Note the usual flow-orthogonal projector $h_{ab}$ becomes 
\begin{align}
h_{ab}\equiv g_{ab}+n_{a}n_{b}=e_{a}e_{b}+N_{ab},
\end{align}
and we can also interpret the 2-metric $N_{ab}$ as being the projector onto
the subspace orthogonal to both $n^{a}$ and $e_{a}$.\end{spacing}} 
the energy-momentum tensor of a perfect fluid is 
\begin{gather}
T_{ab}=\rho n_{a}n_{b}+P(e_{a}e_{b}+N_{ab})\,.\label{eq:pftensor}
\end{gather}
where $\rho$ and $P$ respectively are the energy-density and the
pressure of the fluid as measured by the co-moving observer.\footnote{\vspace{-.3cm}\begin{spacing}{0}In the case when the fluid is also admitting anisotropic stress \citep{Mimoso:2013iga},
the energy momentum tensor reads 
\begin{align}
T_{ab}=\rho n_{a}n_{b}+P(e_{a}e_{b}+N_{ab})+\Pi_{ab}.
\end{align}
In spherical symmetry, the anisotropic stress reduces to $\Pi_{ab}=\Pi\left(N_{ab}-N_{\:c}^{c}e_{a}e_{b}\right)$.
The perfect fluid case then corresponds to $\Pi_{ab}=0$. A cosmological
constant can be modeled as a perfect fluid with $\rho=-P=\frac{\Lambda}{8\pi}$
\citep[see, e.g.][p172]{carroll-2004}.\end{spacing}}

The Einstein Field Equations (EFEs) 
can then be written 
\begin{subequations}\label{eq:system-0} 
\begin{gather}
\dot{E}+
\beta\left(E'+\frac{2(1+E)}{\rho+P(\rho)}\frac{\ud P}{\ud\rho}\rho'\right)=0\,,\label{eq:Eequation-1}\\
\dot{M}_{ms}+
\beta\left(M_{ms}'+4\pi P(\rho)r^{2}\right)=0\,,\label{eq:Mequation-1}\\
M_{ms}'-4\pi r^{2}\,\left(\rho+\frac{\Lambda}{8\pi}\right)=0\,.\label{eq:rhoequation-1}\\
\frac{\alpha'}{\alpha}+\frac{P'}{\rho+P}=0\,.\label{eq:alphaequation-1}
\end{gather}
\end{subequations} where $M_{ms}$ includes both the contribution
of the perfect fluid and that of the cosmological constant, i.e, 
\begin{equation}
M_{ms}=M+\frac{\Lambda}{6}r^{3}\;.
\end{equation}
with $M=4\pi\int_{0}^{r}\,\rho(u',t=t_{0})\,u^{2}\,{\rm d}u$ representing
the total spherical mass-energy of the perfect fluid contained in
a given radius, in spherical symmetry. They are to be complemented
with an equation of state of the form 
\begin{equation}
f(\rho,P)=0\,.\label{eq:stateequation-1}
\end{equation}
Note that the solution, 
with the perfect fluid energy density $\rho$ and isotropic pressure
$P$, can be expressed in terms of $E$, the energy/curvature of the spatial
hypersurfaces, and of the Misner-Sharp mass, $M$, of the fluid. Given that in the
case of multiple fluids, the various corresponding masses, densities
and pressures are summed, and that the energy/curvature results from
the total mass sum, we find for $\beta$ 
\begin{gather}
\beta(t,r)=\pm\alpha(t,r)\sqrt{\frac{2M(t,r)}{r}+\frac{\Lambda r^{2}}{3}+E(t,r)}\,,\label{eq:betaDef}
\end{gather}
where the $-$ sign corresponds to collapse and the $+$ sign, to
expansion\footnote{Note that our definition for $\beta(t,r)$ corresponds to that
in \citet{Lasky:2006zz} with the opposite sign.}.

The system is closed by the following 
data given on the initial $t=t_{0}$ hypersurface: \begin{subequations}\label{eq:initaldata-1}
\begin{gather}
\rho(t_{0},r)=g(r)\,,\label{eq:rho0-1}\\
\dot{\rho}(t_{0},r)=h(r)\,.\label{eq:rhodot0-1}
\end{gather}
\end{subequations}

\noindent In Ref.~\cite{MLeDM09} a separating surface that we 
named 
Matter Trapping Surface (MTS) was defined 
by the following conditions\footnote{\vspace{-.3cm}\begin{spacing}{0}Note that these conditions can be recast in a geometric and gauge
independent form, using the expansion $\Theta_{n}$ of the evolving
2-surface in the MTS along the flow of matter $n^{a}$, that is related
to the flow volume expansion $\Theta_{3}$ and to its spherical shear
scalar $\sigma$ by $\Theta_{n}=2\left(\frac{\Theta_{3}}{3}+\sigma\right)$
. In spherical symmetry, $\frac{\Theta_{3}}{3}+\sigma=\frac{\Lie_{n}r}{r}$.
Therefore, for $r\ne0$, we have equivalence between the geometric
expression of the MTS conditions and the gauge invariant expressions
\begin{align}
\left.\begin{array}{rl}
\Lie_{n}r= & \frac{r}{2}\Theta_{n}=0\\
\Lie_{n}\Lie_{n}r= & \frac{r}{2}\left(\Lie_{n}\Theta_{n}+\frac{\Theta_{n}^{2}}{2}\right)=0
\end{array}\right\} \Leftrightarrow & \left\{ \begin{array}{rl}
\Theta_{n}= & 0,\\
\Lie_{n}\Theta_{n}= & -\frac{\Theta_{n}^{2}}{2}=0\;.
\end{array}\right.
\end{align}
\end{spacing}} 
\begin{subequations} 
\begin{gather}
\Lie_{n}r=0\,,\label{MTS_1}\\
\Lie_{n}\left(\Lie_{n}r\right)=0\,,\label{MTS_2}
\end{gather}
\end{subequations} and hence $\Theta_{n}=\Lie_{n}\Theta_{n}=0$ according
to Eq. (\ref{theta_n}). 

In terms of the integral quantities $E$ and $M_{ms}=M+(\Lambda/6)\,r^{3}$
the MTS conditions read
\begin{subequations} 
\begin{gather}
E=-\frac{2M}{r}-\frac{\Lambda r^{2}}{3}\,,\\
\text{gTOV}\equiv\frac{M}{r^{2}}-\frac{\Lambda r}{3}+4\pi rP+\frac{1+E}{\rho+P}P'=0\,,
\end{gather}
\end{subequations} where gTOV is a functional we proposed in \citet{MLeDM09}
that yields the Tolman-Oppenheimer-Volkoff (TOV) equilibrium equation
\cite{Tolman:1934za,oppenheimer-1939a} on the MTS. The first of
these conditions, sets the locus of turnaround radius \cite{Gunn:1972sv,Fillmore:1984wk,Bertschinger:1985pd,Korkidis:2019nzk,DelPopolo:2020mge},
and was coined the kinematic condition. It reveals that the MTS can
only exist in positively curved/negative energy regions. The second
condition balances the radial acceleration, and may separate regions
of opposite accelerations. It is named the 
dynamic condition, and can be related to the cracking condition first
put forward by Herrera \cite{herrera-1992,diprisco-1994,DiPrisco:1997tw}
in the framework of the gravitational instability of bound spherical
distributions of matter surrounded by vacuum.


The question now that needs clarification is whether the MTS can be
used as a boundary setting that would entirely determine the dynamics
on each side. This would be the case if the EFEs together with the
MTS can be recast as a Cauchy problem.

A Cauchy problem is defined as system of $n$ partial differential
equations in $\mathbb{R}^{m}$, of up to $k$-th order, in $n$ unknown
functions, with boundary/initial data given in an $m-1$ dimensional
hypersurface of $\mathbb{R}^{m}$ for derivatives up to $k-1$-th
order for each of the $n$ unknown functions. Therefore, the system
given by Eqs.~\eqref{eq:system-0} with the set of data \eqref{eq:initaldata-1}
, having unclear choice of unknown functions
, 
is not a Cauchy problem in this form. 
We 
then 
have to rewrite it as an equivalent Cauchy problem in order to use
some mathematical results.



\subsection{The dust and $\Lambda$ model}

\noindent In the case when the perfect fluid source 
reduces to dust and a cosmological constant $\Lambda$, we notice
at once that 
\begin{equation}
\frac{\alpha'}{\alpha}=0\,.\label{dust_alpha_geod}
\end{equation}
From Eq.~(\ref{dust_alpha_geod}), we see that $\alpha$ is only
a function of time and thus we can set $\alpha=1$ by a time rescaling
in the metric \eqref{eq:GPGmetric}. Thus the system of field equations
is reduced to \cite{Lasky:2006zz,MLeDM09,LeDMM09a} 
\begin{subequations}\label{eq:system-1} 
\begin{gather}
\dot{E}+
\beta E'=0\,,\label{eq:dotEdust-1}\\
\dot{M}+
\beta M'=0\,,\label{eq:dotMdust-1}\\
M'-4\pi r^{2}\rho=0\,.\label{eq:MprimeDust}
\end{gather}
\end{subequations} Eqs.~\eqref{eq:dotEdust-1} and \eqref{eq:dotMdust-1}
only explicitly depend on the unknowns $E$, $M$, and inherently
on $\Lambda$, while all the other quantities as for instance $\rho$,
are completely defined by giving these two functions. Therefore, the 
system is reduced to the two former equations. 

The initial data in Eqs.\eqref{eq:initaldata-1} should be rewritten
as a proper set of Cauchy data, that is, initial values for $M(t_{0},r)$
and $E(t_{0},r)$. For $M(t_{0},r)$ we trivially obtain from Eq.~\eqref{eq:MprimeDust}: 
\begin{gather}
M(t_{0},r)=4\pi\int_{0}^{r}\rho(t_{0},s)s^{2}\ud s\,.
\end{gather}

In order to find $E{(}t_{0},r)$, we use $\dot{\rho}(t_{0},r)$ to
write: 
\begin{gather}
\dot{M}_{0}(r)\equiv4\pi\int_{0}^{r}\dot{\rho}(t_{0},s)s^{2}\ud s\,,
\end{gather}
and consistently with Eq.~\eqref{eq:dotMdust-1}, $\dot{M}_{0}(r)=\dot{M}(t_{0},r)$,
implies for $E(t_{0},r)$ (assuming no initial region of dust vacuum,
i.e. $\forall r,M^{\prime}\left(t_{0},r\right)\ne0$), 
from \eqref{eq:betaDef}: 
\begin{gather}
E(t_{0},r)=\left(\frac{\dot{M}_{0}}{M'(t_{0},r)}\right)^{2}-\frac{2M(t_{0},r)}{r}-\frac{\Lambda r^{2}}{3}\,,
\end{gather}
which means that giving $M(t_{0},r),E(t_{0},r)$ is equivalent to
giving $\rho(t_{0},r),\dot{\rho}(t_{0},r)$ at the initial surface.

\noindent The system is now a quasi-linear, first order system, given
by \begin{subequations} \label{eq:Xsystem-1}
\begin{gather}
\dot{X}=-\beta\,X'\,\\
X(t_{0},r)=X_{0}(r)\,,
\end{gather}
\end{subequations} where 
\begin{gather}
X\equiv\left[\begin{array}{c}
E(t,r)\\
M(t,r)
\end{array}\right].\label{eq:CauchyData}
\end{gather}

Since the derivative in the normal direction to the initial Cauchy
surface is $\dot{X}$, which is completely determined in terms of
$X'$ from the initial surface, the system \eqref{eq:Xsystem-1},
together with the Cauchy data \eqref{eq:CauchyData} 
is well posed.


We 
further look for the characteristic curves (CCs) of the system \cite[for
the method of characteristics see][]{courant-hilbert-2}. 
They can be defined as the integral curves of the coordinates for
which 
the system of PDE's is reduced to a system of ODE's of the form 
\begin{gather}
\frac{\ud X}{\ud s}=0.
\end{gather}
This form 
implies 
\begin{gather}
\frac{\partial}{\partial s}=\partial_{t}+\beta\partial_{r}
\end{gather}
which means that $s$ is just the known Lemaitre-Tolman-Bondi proper
time coordinate \cite{Lemaitre:1933gd,Tolman:1934za,bondi-1947,Lasky:2006zz,MLeDM09}.
Note also that $(\partial_{s})^{a}=n^{a}$ which means that the CCs
correspond to the flow lines. Since the equation is homogeneous, the
solution is given by the propagation of the initial values $M(t_{0},r)$
and $E(t_{0},r)$ along the integral curves of $\partial_{s}$. 

At the 
MTS surface separating 
collapse and expansion $\beta=0$, and the CC is locally tangent to
$\partial_{t}$. This condition is equivalent to the turnaround radius
condition $\Lie_{n}r=0$ \cite{MLeDM09}, which means that the areal
radial velocity of the fluid vanishes at that event. In other words
, the surface of this sphere of fluid is instantaneously constant
.

We compute the time derivative of $\beta$: 
\begin{gather}
\dot{\beta}=-\frac{M}{r^{2}}+\frac{\Lambda r}{3}-
\beta\beta'\,,
\end{gather}
which implies that 
\begin{gather}
\frac{\ud\beta}{\ud s}=\Lie_{n}\beta=\dot{\beta}+
\beta\beta'=-\left(\frac{M}{r^{2}}-\frac{\Lambda r}{3}\right)=-\text{gTOV}\,,
\end{gather}
which 
can be interpreted as the areal radial acceleration of a flow line.


We are naturally interested in MTS surfaces where the areal radial
velocity and acceleration vanish. On such a sphere, $\partial_{s}=\partial_{t}$.
Given an initial condition with $\beta(t_{0},r^{*})=0$ and $\dot{\beta}(t_{0},r^{*})=0$
at some $t=t_{0}$ surface $r=r^{*}>0$, as 
the second time derivative reads 
\begin{gather}
\ddot{\beta}=
-%
\beta\left[-\frac{M'}{r^{2}}+\dot{\beta}'\right]
-
\dot{\beta}\beta^{\prime}\,,
\end{gather}
it will vanish on that sphere. This implies that $\beta(t,r^{*})=\dot{\beta}(t,r^{*})=0$
for all $t>t_{0}$, and the sphere will therefore remain unchanged
along the fluid evolution.


This result justifies the definition of matter trapping
shells/surfaces (MTS's), for the dust+$\Lambda$ model.

\subsection{The MTS as the characteristic surface for the dust with $\Lambda$
model}

Consider now the Cauchy problem that consists in solving the system
Eqs.~\eqref{eq:system-1} using the Cauchy surface of the type $r=r^{*}$,
with boundary conditions matching the MTS conditions, instead of the
previous Cauchy surface of the type $t=t_{0}$. 
By the conditions $\beta=\dot{\beta}=0$ for the MTS, we obtain, from
Eqs.~\eqref{eq:dotEdust-1} and \eqref{eq:dotMdust-1}, that the
system only admits solutions if we impose boundary data of the type
\begin{gather}
E(t,r^{*})=E^{*}\,,\quad M(t,r^{*})=M^{*}\,,
\end{gather}
which means that all quantities are constant at that surface.

This leaves us with the following questions
:
\begin{enumerate}
\item Can we integrate inwards and outwards from this surface? 
\item Do we obtain separation between the inner and outer solutions? 
\end{enumerate}
Since this Cauchy surface is a characteristic surface, together with
the system Eq.~\eqref{eq:Xsystem-1}, it constitutes a characteristic
Cauchy problem, and therefore the answer to the first question is
no. This can be seen in detail in the following realization: the 
initial data binds the values of $X$ and $\dot{X}$ at the Cauchy
surface. In order to solve the system we need to compute the derivatives
normal to the Cauchy surface, which in this case means in the $\partial_{r}$
directions. This implies solving the system for $X'$: 
\begin{gather}
-\beta^{-1}\dot{X}=X'\,.
\end{gather}
As, at our Cauchy surface, $\beta=0$, solving the above system is
not possible. This behaviour is typical of characteristic Cauchy problems,
where 
the Cauchy surface coincides with a CC. In such case, the data from
the boundary surface cannot 
propagate outside from it
. Since this is a characteristic Cauchy problem, we have only two
types of solution:

{[}(i){]} 
\begin{enumerate}
\item no solution, if the Cauchy data is not compatible with the problem's
equations. 
\item infinite solutions, if it is 
compatible. In such case, any compatible 
solution of the PDE system can be glued on each side of the Cauchy
surface. \label{ii-1} 
\end{enumerate}
\noindent Solution
~(\ref{ii-1}) is the answer for our second question. In such case,
so long as the Cauchy data remains compatible with the surface values,
the 
interior and exterior solutions may be arbitrary. In other words,
 the MTS Cauchy boundary 
determines neither the interior, nor the 
exterior solution, but it guarantees that they do not depend on each
other. If we impose, on the basis of physical intuition, that the
functions $M$ and $E$ are continuous (which, from a strictly mathematical
viewpoint, is not necessary for a solution to the Cauchy problem
), then 
the MTS can be interpreted in a 
similar way to the continuous matching of the 
static spherical star solutions to the surrounding Schwarzschild spacetime.
The exterior solution only depends on the interior solution 
quantities at the matching sphere, in this case through 
the total (MS) mass and the function $E$, at $r=r*$.



In summary, the MTS in the case of a dust plus $\Lambda$ model acts effectively as a shield that not only bridges different spherically symmetric solutions, but also "protects" each side from the detailed dynamics of the other side. One can think of it as analogous to a thin vacuum \cite[or D-vacuum sphere,][as $\Lambda$ is present]{2016CQGra..33s5006B} spherical shell, that by the virtue of Birkhoff theorem, can be joined to any pair of interior and exterior solutions, provided the continuity conditions required by GR are met.



\section{Illustrative cases with dust and a cosmological constant}
\label{sec:Illustrate}
The conclusion of the previous section reveals that, so long as we have an MTS in a solution to EFEs in spherical symmetry for a  dust and cosmological constant content, it can be matched to any other solution with the same generic conditions. In our previous works, we found such conditions, and in particular the presence of an MTS, in two kinds of solutions: Schwarzschild-de~Sitter (SdS, spherical vacuum with $\Lambda$) and Lema\^itre-Tolman-Bondi-de~Sitter (LTBdS, spherical inhomogeneous dust with $\Lambda$). In order to find the MTS areal radius static shell in a homogeneous Friedman-Lema\^itre-Robertson-Walker-de~Sitter solution, the Einstein static universe must be selected (ES). The general framework for these 3 types of solutions is the LTBdS, and we first recall it.

We then need to characterise the MTS in each of the possible solutions before proposing matching conditions. We start, after the general case, with vacuum, then turn to the ES case, before focusing on the more complex LTBdS. For cosmological likelihood, we choose the outer part of the MTS to be built to asymptote a flat FLRW, while the inner part will follow a classical virialised halo density distribution.

Matching the MTS as a Cauchy surface implies that the PDE variables t and r should remain smooth. This implies the matching MTS should have fixed areal radius. Therefore we will have classes of MTS matchings for each pairs of solutions.
\subsection{MTS in LTB with $\Lambda$}
From our formulation of $\Lambda$CDM LTB in \citet{MLeDM09}, the
governing equations are
\begin{subequations} 
\begin{align}
\beta^2=\dot{r}^{2} & =\frac{2M}{r}+\frac{\Lambda}{3}r^2+E,\label{eq:RadialEvolLTB}\\
\ddot{r} & =-\frac{M}{r^{2}}+\frac{\Lambda}{3}r,\label{eq:RadialAccLTB}
\end{align}
\end{subequations}
with $M$ and $E$ conserved for each shell without shell crossing.
Then, the MTS is defined at
\begin{subequations} 
\begin{gather}
E=-\frac{2M}{r}-\frac{\Lambda r^{2}}{3}\,,\\
\text{gTOV}\equiv\frac{M}{r^{2}}-\frac{\Lambda r}{3}=0\,.
\end{gather}
\end{subequations}
This leads to an MTS for a given $M(r)$ and $E(r)$ profile if both 
\begin{align}
\ddot{r}= & 0:& r_{MTS}= & \sqrt[3]{\frac{3M}{\Lambda}},\label{eq:rMTS}\\
\dot{r}=& 0:&E_{MTS}= & -\left(3M\right)^{\frac{2}{3}}\Lambda^{\frac{1}{3}}.\label{eq:EMTS}
\end{align}
 
\subsection{MTS in Schwarzschild-de~Sitter}

In the case of Schwarzschild-de~Sitter, the Lema\^itre form of the metric reads
\begin{align}
\ud s^{2}= & -\ud T^{2}+\left(\frac{2m}{r}+\frac{\Lambda r^2}{3}\right)\ud R^{2}+r^{2}\ud \Omega^{2}, & r= & r\left(T,R\right),\\
\ud R-\ud T=&\left(\frac{2m}{r}+\frac{\Lambda r^2}{3}\right)^{-\frac{1}{2}}\ud r,&\Rightarrow r\left(T,R\right)=&\left[\sqrt{6m\Lambda}\sinh\left(\frac{3}{2}\sqrt{\frac{\Lambda}{3}}\left(R-T\right)\right)\right]^{\frac{2}{3}}
\end{align}
while the GLTB form, derived from Eq.~\eqref{eq:GPGmetric} for dust with $\Lambda$, adopts that of
\begin{align}
    \beta =&\dot{r},\\
\ud s^{2}= &-\ud t^{2}+\frac{(r^\prime\ud R)^{2}}{1+E(t,R)}+r^{2}\ud\Omega^{2}\,.\label{eq:GLTB}
\end{align}
Therefore identification yields
\begin{align}
    M= & m = cst,\\
    E= & 0.
\end{align}
In this case the MTS in the sense of $\Lambda$LTB can only be found if $\Lambda\to0,r_{MTS}\to\infty$. However, in vacuum, no flow is predefined and we can consider $\dot{r}=0$ everywhere for any $\Lambda$. In this case we can consider that the no acceleration shell $r_{MTS}$ is an MTS.
\subsection{MTS in Einstein-de~Sitter}

In the case of Einstein-de~Sitter, i.e., FLRW with dust and $\Lambda$, we start by writing the line element
\begin{align}
    \ud s^{2}= & -\ud t^{2}+a^{2}\left(\frac{\ud R^{2}}{1-kR^{2}}+R^{2}\ud \Omega^{2}\right), 
\end{align}
matching it with the GLTB form \eqref{eq:GLTB}. Identification in this case yields
\begin{align}
    r= &a R \Rightarrow &r^\prime=& a,\\
    \frac{a^2}{1-kR^{2}}=&\frac{(r^\prime)^{2}}{1+E}=\frac{a^2}{1+E}\Rightarrow & E=&-kR^{2}=-k\left(\frac{r}{a}\right)^{2},\\
    M^\prime =&4\pi r^2 r^\prime \rho(t)\Rightarrow & M=& \frac{4\pi}{3}r^3\rho(t).
\end{align}
The MTS conditions then lead to
\begin{align}
    -k\left(\frac{r}{a}\right)^{2} = & -\left(4\pi \rho(t)\right)^{\frac{2}{3}}\Lambda^{\frac{1}{3}} r^2 \Leftrightarrow k= \left(4\pi a^3\rho(t)\right)^{\frac{2}{3}}\Lambda^{\frac{1}{3}}= cst \Rightarrow \rho(t)=\rho_m(t)\propto a^{-3}, \\
    r_{MTS}= & \sqrt[3]{\frac{4\pi \rho(t)}{\Lambda}}r_{MTS}\Leftrightarrow \rho(t) = \frac{\Lambda}{4\pi}=cst \Rightarrow \Lambda=\Lambda_c=\frac{\kappa}{2}\rho_{m0},a=a_{\star}=1,
\end{align}
so we can only have an MTS in the Einstein static universe, which is filled with MTSs. In that case, $k=\Lambda_c, M=\frac{\Lambda_c}{3}r^3, E=-\Lambda_c r^2$ for any $r$.

\subsection{Building an MTS in LTB example}
In order to obtain an example of MTS separating a virialised structure from a cosmological background, we focus on the building of an NFW core with Hubble central flow going to an outward FLRW-like behaviour.

\subsubsection{The Heaviside MTS jump}
We separate the FLRW denstiy behaviour to the outskirts of the central halo.
Choosing an inner NFW profile, justified by results of Nbody haloes \cite{navarro-1996}, with cosmological background beyond the MTS
\begin{align}
\rho= & \frac{\rho_{0}}{\frac{r}{r_{0}}\left(1+\frac{r}{r_{0}}\right)^{2}}+\rho_{b}\Theta_H\left(r-r_{MTS}\right),
\end{align}
where $\rho_{b}$ is the background energy density recovered for large
values of $r$ beyond the MTS radius, ensured by the Heaviside distribution $\Theta_H$, $\rho_{0}$ defines the energy density scale of the central halo, and 
$r_{0}$ marks the change of density logarithmic slope from the central
cusp $\propto r^{-1}$ to the Keplerian outer decrease as $\propto r^{-3}$. Note that at $r=r_{0},$ $\rho(r_{0})=
\frac{\rho_{0}}{4}$. The corresponding mass then reads 
\begin{align}
M= & 4\pi\left\{ r_{0}^{3}\rho_{0}\left[\ln\left(1+\frac{r}{r_{0}}\right)-\frac{r}{r+r_{0}}\right]+\rho_{b}\frac{r^{3}-r_{MTS}^{3}}{3}\Theta_H\left(r-r_{MTS}\right)\right\} .
\end{align}


At the MTS, the radius is then determined by the scale of $\Lambda$,
inverting $r_{MTS}$ as $M_{MTS}\left(r_{MTS}\right)=\frac{\Lambda}{3}r_{MTS}^{3}$.
This is set by the dynamic condition \eqref{eq:RadialAccLTB} with \eqref{eq:rMTS}.
Using the kinematic condition \eqref{eq:RadialEvolLTB} to ensure that
\eqref{eq:EMTS} is simultaneously verified leads to having vanishing
areal velocity at the MTS.

The dynamic condition determines the position of the MTS by solving
\begin{align}
M\left(r\right)= & \frac{\Lambda}{3}r^{3}\label{eq:MTSmassConstraint}\\
\Leftrightarrow\frac{\left(\frac{\Lambda}{4\pi}-\rho_{b}\Theta_H\right)}{\rho_{0}}\frac{x^{3}}{3} + \frac{\rho_{b}\Theta_H}{\rho_{0}}\frac{x_0^{3}}{3}= & \ln\left(1+x\right)-\frac{x}{x+1},x=\frac{r}{r_{0}},x_0=\frac{r_{MTS}}{r_{0}}.
\end{align}
Choosing at initial time $t=t_{i}$, and considering the flat cosmology
to set the matter and dark energy components, we get 
\begin{align}
\forall t,\Omega_{\Lambda}+\Omega_{m}= & 1 & \frac{\Lambda}{3}= & H_{0}^{2}\Omega_{\Lambda_{0}}=H_{i}^{2}\Omega_{\Lambda_{i}},\\
 &  & \rho_{b_{0}}= & \rho_{b_{i}}a_{i}^{3}=\frac{3H_{0}^{2}}{8\pi}\Omega_{m\,0}\Rightarrow H_{0}^{2}\Omega_{m\,0}=H_{i}^{2}\Omega_{m\,i}a_{i}^{3},\\
\Rightarrow\frac{\Omega_{\Lambda_{0}}}{\Omega_{m\,0}}= & \frac{\Omega_{\Lambda_{i}}}{\left(1-\Omega_{\Lambda_{i}}\right)a_{i}^{3}} & \Rightarrow\Omega_{\Lambda_{i}}= & \frac{a_{i}^{3}\Omega_{\Lambda_{0}}}{\Omega_{m\,0}+a_{i}^{3}\Omega_{\Lambda_{0}}}.
\end{align}In addition, we have the relation
\begin{align}
    \frac{H_{0}^{2}}{H_{i}^{2}}= & \frac{\Omega_{\Lambda_{i}}}{\Omega_{\Lambda_{0}}} = \frac{a_{i}^{3}}{\Omega_{m\,0}+a_{i}^{3}\Omega_{\Lambda_{0}}}.
\end{align}
With the choice of the conditions 
\begin{align}
\rho_{b}\Theta_H= & \rho_{m}\Theta_H\left(r-r_{MTS}\right)=\rho_{c\;i}\Omega_{mi}\Theta_H\left(r-r_{MTS}\right)=0&r<&r_{MTS}
,\\
\frac{\Lambda}{4\pi}= & 2\rho_{\Lambda}=2\rho_{c\,i}\Omega_{\Lambda i},
\end{align}
the constraint \eqref{eq:MTSmassConstraint} 
constraint 
then becomes 
\begin{align}
\frac{\rho_{c}}{\rho_{0}}\left(2\Omega_{\Lambda}\right)\frac{x^{3}}{3}=&\frac{\rho_{c}}{\rho_{0}}\frac{2 a_{i}^{3}\Omega_{\Lambda_{0}}}{\Omega_{m\,0}+a_{i}^{3}\Omega_{\Lambda_{0}}}\frac{x^{3}}{3}=\ln\left(1+x\right)-\frac{x}{x+1},
\end{align}
which admits one nonzero solution. For example, choosing 
\begin{align}
\Omega_{\Lambda_{0}}= & 0.7, & \Omega_{m_{0}}= & 0.3,& a_{i}= & 10^{-2}, & \rho_{0}= & 200\rho_{c},
\end{align}
the numerical solution to
\begin{align}
\frac{a_{i}^{3}\Omega_{\Lambda_{0}}}{\Omega_{m\,0}}-\frac{3\rho_{0}}{2\rho_{c}x^{3}}\left(\ln\left(1+x\right)-\frac{x}{x+1}\right)\simeq & 0,
\end{align}
yields $\frac{r_{MTS}}{r_{0}}=x_{0}\simeq907.5$.

\subsubsection{Smooth cosmological transition}
In fact we may want a smoother cut off than the Heaviside distribution. 

\paragraph{Induced initial time limit}

The small, finite value of the Heaviside replacement at the MTS imposes that the initial time occurs after a cutoff time as deduced below in a simple approximation.

If we still want the above approximation to be correct at initial time $a_i$, we can impose a condition on the value of the smooth function modeled as a constant $T_H$, as
\begin{align}
    \frac{2\Omega_{\Lambda i}}{T_H\Omega_{m\,i}} =& \frac{2}{T_H}\frac{\Omega_{\Lambda 0}}{\Omega_{m\,0}}a_{i}^{3}\simeq 10^{h}\\
    \Leftrightarrow T_H=&\frac{2\Omega_{\Lambda 0}}{\Omega_{m\,0}}a_{i}^{3}10^{-h} =\frac{14}{3}10^{-h-6}
\end{align}then the existence of a solution to the constraint \eqref{eq:MTSmassConstraint} is leading to 
\begin{align}
    2\rho_{\Lambda}-\rho_{b}=& \rho_{c\,i}\left(2\Omega_{\Lambda i}-T_H\Omega_{m\,i}\right)>0\\
    \Leftrightarrow 2\Omega_{\Lambda i}-T_H\Omega_{m\,i}=&\frac{2 a_{i}^{3}\Omega_{\Lambda_{0}}-T_H\Omega_{m\,0}\left(\Omega_{m\,0}a_{i}^{-3}+\Omega_{\Lambda_{0}}\right)}{\Omega_{m\,0}+a_{i}^{3}\Omega_{\Lambda_{0}}}>0\\
    \Leftrightarrow a_{i}^{6}- T_H\frac{\Omega_{m\,0}}{2}a_{i}^{3}- &  T_H\frac{\Omega_{m\,0}^2}{2\Omega_{\Lambda_{0}}}=\left(a_{i}^{3}-T_H\frac{\Omega_{m\,0}}{4}\right)^2 - T_H\frac{\Omega_{m\,0}^2}{2\Omega_{\Lambda_{0}}} - \left(T_H\frac{\Omega_{m\,0}}{4}\right)^2>0\\
    \Rightarrow a_{i}^{3} > & \sqrt{T_H}\left(\sqrt{T_H}\frac{\Omega_{m\,0}}{4} + \sqrt{\frac{\Omega_{m\,0}^2}{2\Omega_{\Lambda_{0}}} + T_H\left(\frac{\Omega_{m\,0}}{4}\right)^2}\right)=a_{i \rm{min}}^3\simeq\Omega_{m\,0}\sqrt{\frac{T_H}{2\Omega_{\Lambda_{0}}}}\\
    \Rightarrow a_{i}> & \left(\Omega_{m\,0}\sqrt{\frac{T_H}{2\Omega_{\Lambda_{0}}}}\right)^{\frac{1}{3}}=\sqrt{a_{i}}10^{-\frac{h}{6}}\Omega_{m\,0}^{\frac{1}{6}}\simeq a_{i \rm{min}}\\
    \Rightarrow h >&6-\log\left(\Omega_{m\,0}\right) \simeq 6.52. 
\end{align}
If we choose $h=7$, we get
\begin{align}
    T_H=&4.67\times 10^{-13}\\
    a_{i \rm{min}}\simeq& 8.33\times 10^{-3},
\end{align}
so the MTS only exists after some evolution of the background cosmology. In fact, as $\rho_b$ decreases with cosmological evolution, the real cutoff is probably closer to our choice of $a_i$.

\paragraph{Smoother function}

\begin{figure}
\subfloat[\label{fig:FermiFunction}Plot of the Fermi function for the chosen parameters (see text).]{
\includegraphics[width=.45\columnwidth]{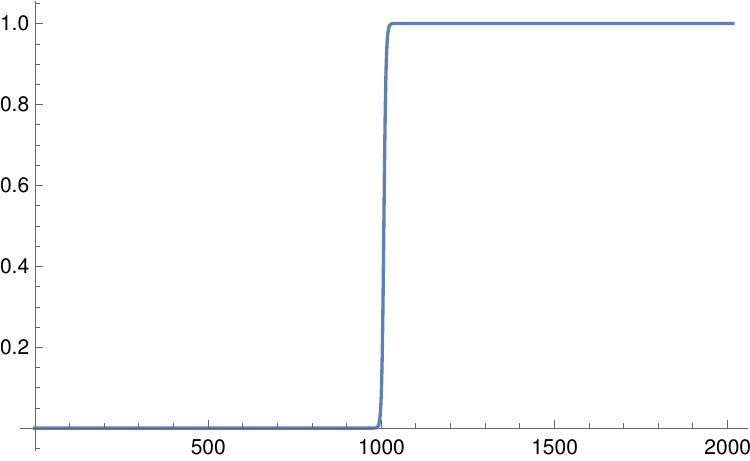}}
\subfloat[\label{fig:DensityFermi}Plot of the density factor for the background density corresponding to the Fermi function mass profile.]{
\includegraphics[width=.55\columnwidth]{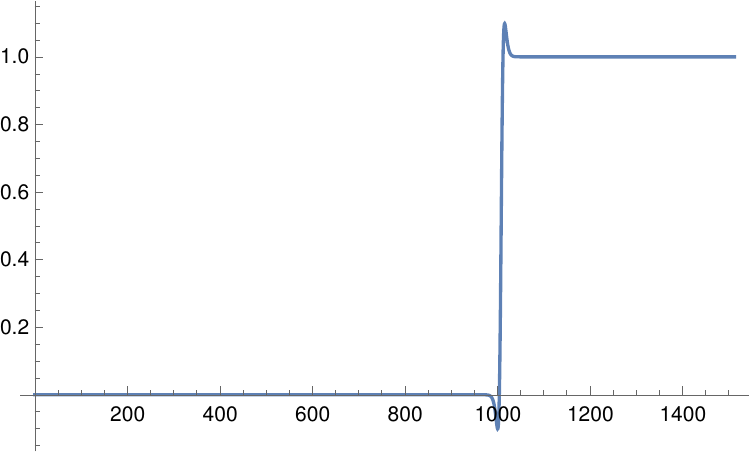}}\\
\subfloat[\label{fig:TotalMass}Total mass profile]{
\includegraphics[width=.5\columnwidth]{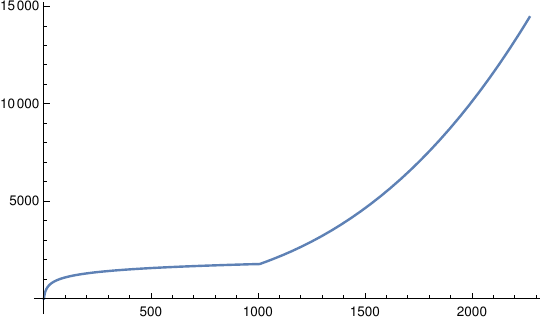}}

\caption{Mass determination with smooth transition}

\end{figure}

If we want to use a smoother function instead of $\Theta_H$, we can choose a type of sigmoid function, namely the Fermi function, for the mass and derive it to get the density. We choose to dissociate the MTS from the transition location at $x_T$ so as to get a controlled, small value of the function and use the above approximation to ensure a similar cutoff time. 
\begin{align*}
\Theta_{H}\left(x-x_{T}\right)\textrm{ in }M&\rightarrow& F_{H}\left(x=\frac{r}{r_{0}},k\right)= & \frac{1}{1+e^{-k\left(x-x_{T}\right)}},\\
\Theta_{H}\left(x-x_{T}\right)\textrm{ in }\rho&\rightarrow &f_{H}\left(x\right)=\frac{\left[\left(x^{3}-x_{T}^{3}\right)F_{H}\right]^{\prime}}{3x^{2}}= & F_{H}\left(x,k\right)\left[1+k\frac{\left(x^{3}-x_{T}^{3}\right)}{3x^{2}}e^{-k\left(x-x_{T}\right)}F_{H}\left(x,k\right)\right].
\end{align*}
They are plotted in Figs.~\ref{fig:FermiFunction} and \ref{fig:DensityFermi}, 
for the choice $x_T=\frac{x_0}{\alpha}$, $\alpha=0.9$ and $k=0.32$. This choice ensures that $\left|F_H\left(x\right)\right|\underset{x\lesssim x_{0}}{\lesssim}9.70\times 10^{-15}<T_H$ for $x<x_0$, so the approximations for the determination of $x_0$ are secured.
The new mass, plotted in Fig.~\ref{fig:TotalMass}, and density functions then read, using the previous prescriptions, 
\begin{align}
M= & H_{i}^{2}r_{0}^{3}\left\{ 300\left[\ln\left(1+x\right)-\frac{x}{1+x}\right]+\frac{H_{0}^{2}}{2H_{i}^{2}}\Omega_{m\,0}a_{i}^{-3}\left(x^{3}-x_{T}^{3}\right)F_{H}\left(x\right)\right\} \nonumber\\
= & H_{i}^{2}r_{0}^{3}\left\{ 300\left[\ln\left(1+x\right)-\frac{x}{1+x}\right]+\frac{1}{2}\frac{\Omega_{m\,0}}{\Omega_{m\,0}+a_{i}^{3}\Omega_{\Lambda_{0}}}\left(x^{3}-x_{T}^{3}\right)F_{H}\left(x\right)\right\} ,\label{eq:TotalMassProfile}\\
\rho= & \frac{M^{\prime}}{4\pi r_{0}^{2}x^{2}}=\frac{H_{i}^{2}r_{0}}{4\pi}\left\{ 300\left[\frac{1}{x\left(1+x\right)^{2}}\right]+\frac{3H_{0}^{2}}{2H_{i}^{2}}\Omega_{m\,0}a_{i}^{-3}f_{H}\left(x\right)\right\} \nonumber\\
= & \frac{H_{i}^{2}r_{0}}{4\pi}\left\{ 300\left[\frac{1}{x\left(1+x\right)^{2}}\right]+\frac{3}{2}\frac{\Omega_{m\,0}}{\Omega_{m\,0}+a_{i}^{3}\Omega_{\Lambda_{0}}}f_{H}\left(x\right)\right\} .\label{eq:DensityLTBdS}
\end{align}

\subsubsection{Kinematic profile condition}
The kinematic condition just requires an areal velocity profile that
ensures $\dot{r}=0$ at the MTS. We opt for a Hubble behaviour near
the centre and at infinity to meet the standard cosmological framework.

Note that the Hubble parameter value need not be the same on each
side of the MTS as they both can yield the $\dot{r}=0$ condition.
Samely the mass profile inside and outside the MTS can differ so long
as they yield the same density and mass at the MTS.

The MTS condition suggests an $\left(r-r_{MTS}\right)^{2}$ factor
to ensure both MTS condition and infinity Hubble-type velocity. An
additional factor of $\tanh^{2}\left(\frac{r}{r_{MTS}}\right)$ ensures
Hubble behaviour in the centre. Defining $H_{i\,\textrm{in}}^{2}=H_{i}^{2}q$ and $H_{i\,\textrm{out}}^{2}=H_{i}^{2}$, we can then propose the areal velocity
profile 
\begin{align}
\dot{r}^{2}= & H_{i}^{2}\left(q\Theta_{H}\left(x_{0}-x\right)+\Theta_{H}\left(x-x_{0}\right)\right)r_{0}^{2}\tanh^{2}\left(\frac{x}{x_{0}}\right)\left(x-x_{0}\right)^{2}.
\end{align}
As the MTS setting is independent of the Hubble parameter value, we can again approximate the shifted Heaviside distribution with $F_{H}\left(x,k\right)$, maintaining the position of the MTS at $x_0$ so the velocity profile can be written
\begin{align}
    \dot{r}^{2}= & H_{i}^{2}\left(qF_{H}\left(x,-k\right)+F_{H}\left(x,k\right)\right)r_{0}^{2}\tanh^{2}\left(\frac{x}{x_{0}}\right)\left(x-x_{0}\right)^{2}.\label{eq:TotalVelocity}
\end{align}
The choice of $q$ follows from ensuring the $E$ profile \eqref{eq:RadialEvolLTB} only intersects the $E_{MTS}$ profile \eqref{eq:EMTS} at the MTS (and possibly at the origin), leading to inner trapped shells below the MTS so no shell crossing can occur \cite{MLeDM09}, as opposed to \cite{LeDMM09a}. Noting $e\equiv\frac{E}{x^{2}r_{0}^{2}H_{i}^{2}},m\equiv\frac{M}{x^{3}r_{0}^{3}H_{i}^{2}},\dot{\chi}^{2}\equiv\frac{\dot{x}^{2}}{H_{i}^{2}x^{2}},\frac{\Lambda}{3}=H_{i}^{2}\Omega_{\Lambda_{i}}$
and $e_{M}\equiv\frac{E_{MTS}}{x^{2}r_{0}^{2}H_{i}^{2}}$, Eqs.~\eqref{eq:RadialEvolLTB} and \eqref{eq:EMTS} lead to
\begin{subequations}
    \begin{align}
e= & \dot{\chi}^{2}-2m-\Omega_{\Lambda_{i}},\\
e_{M}= & -3m^{\frac{2}{3}}\Omega_{\Lambda_{i}}^{\frac{1}{3}}.
\end{align}
\end{subequations}
\begin{figure}
\subfloat[\label{fig:Adjusting-inner-Hubble}Balancing the kinetic term with $\Delta e_{W}$. The blue curve represents $\Delta e_{W}$, while the brown curve shows $K$, with notations and parameter choices indicated in the text.]{\includegraphics[width=0.5\columnwidth]{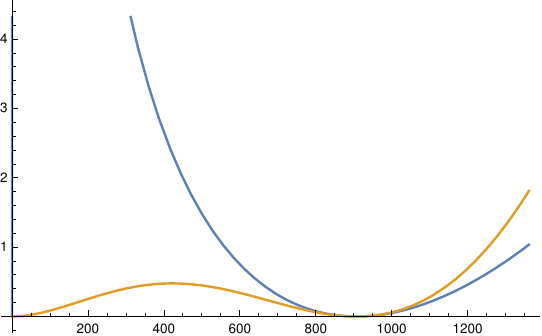}
}\subfloat[\label{fig:Resulting-intersection}Resulting intersection between $E$ and $E_{MTS}$. The blue curve represents $E$, while the brown curve shows $E_{MTS}$, with notations and parameter choices indicated in the text.the range focusses on the intersection.]{\includegraphics[width=0.5\columnwidth]{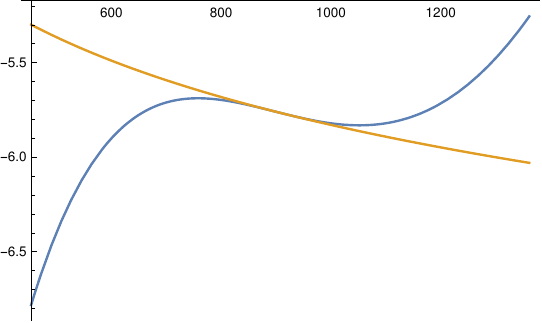}

}

\caption{Adjusting inner Hubble parameter
to the no-shell-crossing condition.}

\end{figure}
Thus the constraint $E<E_{MTS}$ for $x<x_{0}$, with the values
\begin{subequations}
    \begin{align}
m= & \frac{300}{x^{3}}\left[\ln\left(1+x\right)-\frac{x}{1+x}\right]+\frac{1}{2}\frac{\Omega_{m\,0}}{\Omega_{m\,0}+a_{i}^{3}\Omega_{\Lambda_{0}}}\left(1-\left(\frac{x_{T}}{x}\right)^{3}\right)F_{H}\left(x\right)\nonumber \\
& \underset{x<x_{0}}{\simeq}\frac{300}{x^{3}}\left[\ln\left(1+x\right)-\frac{x}{1+x}\right],\\
\dot{\chi}^{2}= & \left(qF_{H}\left(x,-k\right)+F_{H}\left(x,k\right)\right)\tanh^{2}\left(\frac{x}{x_{0}}\right)\left(1-\frac{x_{0}}{x}\right)^{2}\underset{x<x_{0}}{\simeq}q\tanh^{2}\left(\frac{x}{x_{0}}\right)\left(1-\frac{x_{0}}{x}\right)^{2},
\end{align}
\end{subequations}
leads to
\begin{align}
\frac{E_{MTS}-E}{x^{2}r_{0}^{2}H_{i}^{2}}>0\Leftrightarrow & \Delta e_{W}\left(x\right)=2m+\Omega_{\Lambda_{i}}-3m^{\frac{2}{3}}\Omega_{\Lambda_{i}}^{\frac{1}{3}}>\dot{\chi}^{2}=K\left(x\right)\ge0.
\end{align}
Plotting $\Delta e_{W}\left(x\right)$ reveals that it is strictly positive, except at $x_0$. This restricts the interior Hubble parameter, as $\Delta e_{W}$ is tangent to the horizontal at $x_0$, since
\begin{figure}
\hspace{-3.9cm}\begin{minipage}[t]{1\paperwidth}\hspace{-3.6cm}\subfloat[$E$ vs $E_{MTS}$ profiles for the model with radius ranging from
0 to 2.5$x_{0}$, showing restricted total vertical range to illustrate
the global behaviour of $E$.]{\includegraphics[width=0.45\columnwidth]{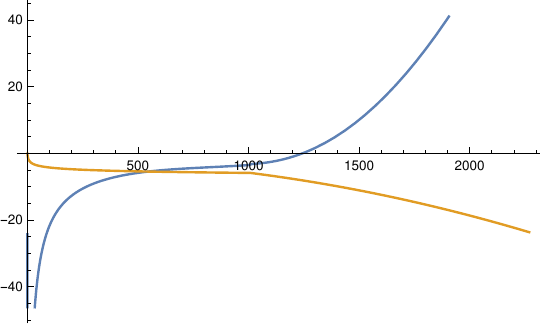}
}
\subfloat[$E$ vs $E_{MTS}$ profiles for the model with radius ranging from
0 to 2.5$x_{0}$, showing total vertical range to illustrate the central
behaviour of $E$.]{\includegraphics[width=0.45\columnwidth]{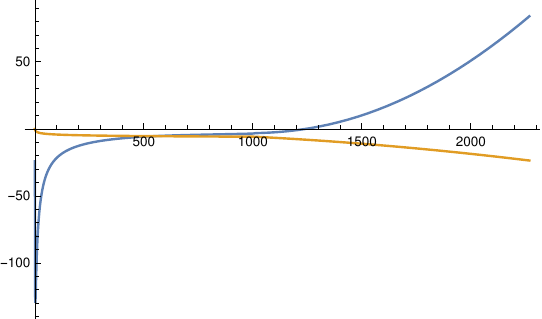}
}\end{minipage}

\subfloat[$E$ vs $E_{MTS}$ profiles for the model in linear-log scales, showing
restricted ranges to illustrate the intersection of $E$ with $E_{MTS}$.]{\includegraphics[width=0.5\columnwidth]{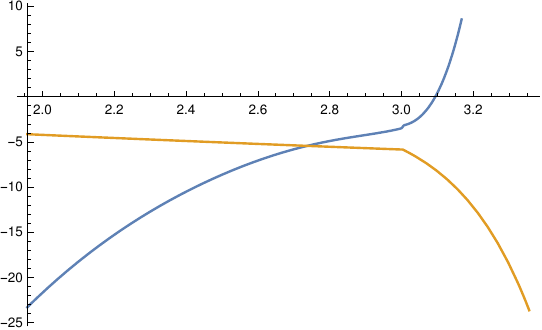}
}

\caption{GLTB model with NFW inner halo and FLRW outer behaviour containing
a stable MTS. The velocity jump is restricted to a factor 5 for illustration.}
\label{GLTB-model}
\end{figure}
\begin{align}
\Delta e_{W}\left(x_{0}\right)= & 0, & \Delta e_{W}^{\prime}\left(x_{0}\right)= & 0, & \Delta e_{W}^{\prime\prime}\left(x_{0}\right)= & 6\Omega_{\Lambda_{i}}\left(\frac{100}{x_{0}^{2}\left(1+x_{0}\right)^{2}}-\frac{\Omega_{\Lambda_{i}}}{x_{0}}\right)^{2}\simeq1.51\times10^{-11},
\end{align}
while the kinetic term yields
\begin{align}
K\left(x_{0}\right)= & 0, & K^{\prime}\left(x_{0}\right)= & 0, & K^{\prime\prime}\left(x_{0}\right)= & 2q\left(\frac{\tanh\left(1\right)}{x_{0}}\right)^{2}\simeq1.41\times10^{-6}q.
\end{align}
In the range $x<x_{0}$ in this case, to ensure no intersection before $x_0$, we can choose \begin{minipage}[t]{2.1cm}%
$\frac{K^{\prime\prime}\left(x_{0}\right)}{\Delta e_{W}^{\prime\prime}\left(x_{0}\right)}\le1$
\end{minipage} 
$\Rightarrow q\le q_{max}\simeq1.07\times10^{-5}$. The $q_{max}$ choice avoids further shell crossings outside of $x_0$, so we adopt it in Fig.~\ref{fig:Adjusting-inner-Hubble}.
This approximate no-shell-crossing condition leads to $E<E_{MTS}$ for $x<x_0$, while $E>E_{MTS}$ for $x>x_0$, as illustrated in Fig.~\ref{fig:Resulting-intersection}
Then the corresponding $E$ and $E_{MTS}$ profiles are obtained.
\begin{table}
\hspace*{-.5cm}%
\noindent\begin{minipage}[t]{1\linewidth}%
\begin{center}
\begin{tabular}{>{\centering\arraybackslash}m{2.5cm}|>{\centering\arraybackslash}m{2.2cm}
>{\centering\arraybackslash}m{3.4cm}
>{\centering\arraybackslash}m{3cm}
>{\centering\arraybackslash}m{2.3cm}
>{\centering\arraybackslash}m{2.8cm}}
observable & $\rho$ & $M$ & $E$ & $\Lambda$ & $r_{MTS}$\tabularnewline
\hline
SdS & 0 & $m$ & 0 & $\Lambda$ & $\sqrt[3]{\frac{3m}{\Lambda}}$\tabularnewline
\hline
ES & $\frac{M^{\prime}}{4\pi r^{2}r^{\prime}}=\rho_{m0}$ & $\frac{\Lambda_{c}}{3}r^{3}$ & $-\Lambda_{c}r^{2}$ & $\Lambda_{c}=4\pi\rho_{m0}$ & $\forall r\Rightarrow r=\sqrt[3]{\frac{3m}{\Lambda}}$\tabularnewline
\hline
SdS to ES jump at $r_{MTS}$ & $\Delta\rho=\rho_{m0}$ & $\Delta M=\left(\frac{\Lambda_{c}}{\Lambda}-1\right)m$ & $\Delta E=-\Lambda_{c}r_{MTS}^{2}$ & $\Delta\Lambda=\Lambda_{c}-\Lambda$ & 0\tabularnewline
\end{tabular}
\par\end{center}%
\end{minipage}

\caption{\label{tab:SdS--EdS-juntion}SdS--ES junction}

\end{table}

\subsubsection{Total $E$ and $E_{MTS}$ profiles}

The GLTB model is then determined by the total mass and velocity profiles \eqref{eq:TotalMassProfile} and \eqref{eq:TotalVelocity}, inputted in $E$ and $E_{MTS}$ in Eq.~\eqref{eq:EMTS} and
\begin{align}
E= & H_{i}^{2}r_{0}^{2}\left(\left(qF_{H}\left(x,-k\right)+F_{H}\left(x,k\right)\right)\tanh^{2}\left(\frac{x}{x_{0}}\right)\left(x-x_{0}\right)^{2}-\frac{600}{x}\left[\ln\left(1+x\right)-\frac{x}{1+x}\right]\vphantom{\frac{\Omega_{m\,0}}{\Omega_{m\,0}+a_{i}^{3}\Omega_{\Lambda_{0}}}}\right.\nonumber \\
 & \left.-\frac{\Omega_{m\,0}}{\Omega_{m\,0}+a_{i}^{3}\Omega_{\Lambda_{0}}}x^{2}-\frac{\Omega_{m\,0}}{\Omega_{m\,0}+a_{i}^{3}\Omega_{\Lambda_{0}}}\left(x^{3}-x_{T}^{3}\right)\frac{F_{H}\left(x\right)}{x}x^{2}\right).\label{eq:E_LTBdS}
\end{align}
Given the value of $q$ defined above, the jump in velocity makes it numerically overwhelming passed the MTS so we represent the total profiles with a jump reduced to a factor 5 for illustration purpose. This allows for a representation of the total model In Fig.~\ref{GLTB-model}. 

\subsection{MTS as Matching surface between different spacetimes}

Armed with the models above, we can produce matching between the three different models studied, Schwarzschild-de~Sitter (SdS),  Einstein-static (ES) and $\Lambda$LTB (LTBdS), generating 6 different models designated as "inner spacetime"--"outer spacetime": SdS--ES, SdS--LTBdS, ES--SdS, ES--LTBdS, LTBdS--SdS, and LTBdS--ES.

\subsubsection{SdS--ES and ES--SdS}
In the first case we have a vacuum sphere surrounded by a static dust, while the second case surrounds a central static dust ball by vacuum. Separation is made at the MTS for the same continuous $r=r_{MTS}$. However the mass, curvature and cosmological constant need not be continuous at the MTS. This is illustrated in Table~\ref{tab:SdS--EdS-juntion}.

\subsubsection{SdS--LTBdS and LTBdS--SdS}
In the first case we have a vacuum sphere surrounded by an expanding dust LTBdS, while the second presents the collapsing core of the LTBdS surrounded by vacuum. The continuous radial separation is determined this time by the static MTS of the LTBdS side, at $r_{MTS}=x_0 r_0$. Again, the mass, curvature and cosmological constant need not be continuous, as shown in Table~\ref{tab:SdS--LTBdS-juntion}.
\begin{table}
\hspace*{-5cm}%
\noindent\begin{minipage}[t]{1\linewidth}%
\begin{center}
\begin{tabular}{>{\centering}m{2.8cm}|ccccc}
observable & $\rho$ & $M$ & $E$ & $\Lambda$ & $r_{MTS}$\tabularnewline
\hline 
SdS & 0 & $m$ & 0 & $\Lambda_{SdS}$ & $\sqrt[3]{\frac{3m}{\Lambda_{SdS}}}$\tabularnewline
\hline 
LTBdS & $\rho=$Eq.~\eqref{eq:DensityLTBdS} & $M=$Eq.~\eqref{eq:TotalMassProfile} & $E=$Eq.~\eqref{eq:E_LTBdS} & $\Lambda=3H_{i}^{2}\Omega_{\Lambda_{i}}$ & $x_{0}r_{0}\Rightarrow r_{0}=\sqrt[3]{\frac{3m}{\Lambda_{SdS}x_{0}^{3}}}$\tabularnewline
\hline 
SdS to LTBdS jump at $r_{MTS}$ & $\Delta\rho=\rho_{m0}$ & $\Delta M=\frac{\Lambda x_{0}}{\sqrt[3]{3m^{2}\Lambda_{SdS}^{2}}}-m$ & $\Delta E=-H_{i}^{2}\sqrt[3]{\frac{9m^2}{\Lambda_{SdS}^{2}}}$ & $\Delta\Lambda=3H_{i}^{2}\Omega_{\Lambda_{i}}-\Lambda_{SdS}$ & 0\tabularnewline
\end{tabular}
\par\end{center}%
\end{minipage}

\caption{\label{tab:SdS--LTBdS-juntion}SdS--LTBdS junction}
\end{table}

\subsubsection{ES--LTBdS and LTBdS--ES}

In the first case, a central static dust ball is surrounded by an expanding dust LTBdS, while the second illustrates a collapsing core of the LTBdS surrounded by a static dust environment. The continuous spatial separation at the MTS is set by the choice of NFW scale $r_0$. As in previous cases, the mass, curvature and cosmological constant need not be continuous, as shown in Table~\ref{tab:EdS--LTBdS-juntion}.

\begin{table}
\hspace*{-3cm}%
\noindent\begin{minipage}[t]{1\linewidth}%
\begin{center}
\begin{tabular}{>{\centering}m{2.8cm}|ccccc}
observable & $\rho$ & $M$ & $E$ & $\Lambda$ & $r_{MTS}$\tabularnewline
\hline 
ES & $\frac{M^{\prime}}{4\pi r^{2}r^{\prime}}=\rho_{m0}$ & $\frac{\Lambda_{c}}{3}r^{3}$ & $-\Lambda_{c}r^{2}$ & $\Lambda_{c}=4\pi\rho_{m0}$ & $\forall r$\tabularnewline
\hline 
LTBdS & $\rho=$Eq.~\eqref{eq:DensityLTBdS} & $M=$Eq.~\eqref{eq:TotalMassProfile} & $E=$Eq.~\eqref{eq:E_LTBdS} & $\Lambda=3H_{i}^{2}\Omega_{\Lambda_{i}}$ & $x_{0}r_{0}$\tabularnewline
\hline 
ES to LTBdS jump at $r_{MTS}$ & $\Delta\rho=0$ & $\Delta M=\left(\frac{\Lambda}{r_{0}}-\Lambda_{c}\right)\frac{r_{MTS}^{3}}{3}$ & $\Delta E=\left(\Lambda_{c}-H_{i}^{2}\right)r_{MTS}^{2}$ & $\Delta\Lambda=3H_{i}^{2}\Omega_{\Lambda_{i}}-4\pi\rho_{m0}$ & 0\tabularnewline
\end{tabular}
\par\end{center}%
\end{minipage}

\caption{\label{tab:EdS--LTBdS-juntion} ES--LTBdS junction}
\end{table}

\section{Conclusion}

\label{sec:conclusion}



In this paper, we have rigorously examined Matter Trapping Surfaces (MTS) within the framework of cosmology and gravitation by formulating the problem as a Cauchy problem for dust with a cosmological constant ($\Lambda$). Our main result is 
the establishment 
that MTSs are characteristic surfaces of the Cauchy problem, generated by the characteristic curves of the 
PDE system
. This implies two possible outcomes: either there is no solution to the Cauchy problem, or there are infinite solutions. In the latter case, any solution to the partial differential equation (PDE) system on one side of the MTS can be extended to either side of the MTS.

To illustrate 
the effects of this proposition
, we presented three examples containing MTSs—Einstein-static (ES), Schwarzschild-de Sitter (SdS), and Lemaître-Tolman-Bondi-de Sitter (LTBdS) models—forming six combinations of solutions. Specifically, we developed an LTBdS model with a central Navarro-Frenk-White (NFW) profile and a Heaviside-limited FLRW-like expanding outer region. These examples demonstrate the practical application of 
the mathematical properties we found 
and show that the LTBdS model can present a static, stable MTS for the first time.

The implications of these findings are significant. By showing that MTSs can be used to glue different solutions on either side, our work provides a powerful tool for constructing and analyzing complex cosmological models. This approach not only advances theoretical understanding but also offers a robust method for modeling matter distribution and dynamics in the universe.

In addition, this result strengthens the interpretation of the MTS as the separating shell between expanding and collapsing regions. For a $\Lambda$CDM matter model the separation is even sharper, since the MTS acts as a boundary beyond which no detailed information about the dynamic flow in the interior propagates to the exterior, and vice versa. This makes it possible to match any interior and exterior solutions that satisfy the minimal physical conditions required to be interpreted as a spacetime. In this sense, the result is reminiscent of the situation in which, by virtue of Birkhoff’s theorem, one may join any interior Schwarzschild solution to the exterior vacuum Schwarzschild spacetime provided they share the same mass and areal radius at the junction surface.

Following these results, a natural question is whether this MTS separation effect also persists in models with non-zero pressure, where the dynamics become substantially more intricate. It is also worth emphasizing that the present analysis assumes strict spherical symmetry. Therefore, a natural extension would be to incorporate angular momentum in the dust shells and to explore alternative symmetry classes.

In conclusion, our study establishes a solid foundation for the use of MTSs in cosmology and gravitation, offering new pathways for research and practical applications. The ability to generate and utilize stable MTSs in various cosmological scenarios opens up significant opportunities for advancing our understanding of the universe's structure and evolution.


\begin{acknowledgements}
    MLeD acknowledges the financial support by the Lanzhou University starting fund, the Fundamental Research Funds for the Central Universities (Grants No. lzujbky-2019-25 and lzujbky-2025-jdzx07), the Natural Science Foundation of Gansu Province (No. 22JR5RA389 and No.25JRRA799), National Science Foundation of China  (NSFC grant No.12247101)
and the ‘111 Center’ under Grant No. B20063.

\end{acknowledgements}
\bibliographystyle{apsrev4-1}
\bibliography{shortnames,referenciasMerge}

@String { arXiv   = {http://www.arxiv.org/abs/} }

@Unpublished{coxa,
  Title                    = {COXA},
  Author                   = {{\huge{COXA}}}
}

@Article{adler-2005,
  Title   = {Simple Analytical Models of Gravitational Collapse},
  Author  = {R. J. Adler and J. D. Bjorken and P. Chen and J. S. Liu},
  Journal = {American Journal of Physics},
  Year    = {2005},
  Pages   = {1148--1159},
  Volume  = {73},
  Issue   = {12},
  Doi     = {10.1119/1.2117187},
  Eprint  = {gr-qc/0502040},
  ArchivePrefix = {arXiv},
}

@Article{Arnowitt:1959ah,
  Title                    = {Dynamical Structure and Definition of Energy in General Relativity},
  Author                   = {Richard L. Arnowitt and Stanley Deser and Charles W. Misner},
  Journal                  = physrev,
  Year                     = {1959},
  Pages                    = {1322--1330},
  Volume                   = {116},

  Doi                      = {10.1103/PhysRev.116.1322}
}

@Article{bondi-1947,
  Title                    = {Spherically Symmetrical Models in General Relativity},
  Author                   = {H. Bondi},
  Journal                  = mn,
  Year                     = {1947},
  Pages                    = {410--425},
  Volume                   = {107},
  Url                      = {http://adsabs.harvard.edu/abs/1947MNRAS.107..410B}
}

@Article{2016CQGra..33s5006B,
  Title                    = {{The Birkhoff theorem and string clouds}},
  Author                   = {{Bronnikov}, K.~A. and {Kim}, S.-W. and {Skvortsova}, M.~V.},
  Journal                  = {Classical and Quantum Gravity},
  Year                     = {2016},

  Month                    = oct,
  Number                   = {19},
  Pages                    = {195006},
  Volume                   = {33},
  Adsurl                   = {http://adsabs.harvard.edu/abs/2016CQGra..33s5006B},
  Archiveprefix            = {arXiv},
  Doi                      = {10.1088/0264-9381/33/19/195006},
  Eid                      = {195006},
  Eprint                   = {1604.04905},
  Primaryclass             = {gr-qc}
}

@Book{carroll-2004,
  Title                    = {Spacetime and Geometry},
  Author                   = {Sean Carroll},
  Publisher                = {Addison-Wesley Publishing Company},
  Year                     = {2004},

  Address                  = {San Francisco}
}

@Book{courant-hilbert-2,
  Title                    = {Methods of Mathematical Physics},
  Author                   = {Richard Courant and David Hilbert},
  Publisher                = {Wiley-VCH},
  Year                     = {1962},
  Volume                   = {2}
}

@Article{diprisco-1994,
  Title                    = {Tidal forces and fragmentation of self-gravitating compact objects},
  Author                   = {{Di Prisco}, A. and E. Fuenmayor and L. Herrera and V. Varela},
  Journal                  = pla,
  Year                     = {1994},
  Number                   = {1},
  Pages                    = {23--26},
  Volume                   = {195},

  Doi                      = {10.1016/0375-9601(94)90420-0},
  ISSN                     = {0375-9601}
}

@Article{herrera-1992,
  Title                    = {Cracking of self-gravitating compact objects},
  Author                   = {L. Herrera},
  Journal                  = pla,
  Year                     = {1992},
  Number                   = {3},
  Pages                    = {206--210},
  Volume                   = {165},

  Doi                      = {10.1016/0375-9601(92)90036-L},
  ISSN                     = {0375-9601}
}

@Article{Lasky:2006zz,
  Title                    = {Spherically Symmetric Gravitational Collapse of General Fluids},
  Author                   = {Paul D. Lasky and Anthony W. C. Lun},
  Journal                  = prd,
  Year                     = {2007},
  Pages                    = {024031},
  Volume                   = {75},

  Archiveprefix            = {arXiv},
  Doi                      = {10.1103/PhysRevD.75.024031},
  Eprint                   = {gr-qc/0612007},
  Primaryclass             = {gr-qc}
}

@Article{LaskyLun07,
  Title                    = {Gravitational collapse of spherically symmetric plasmas in {Einstein-Maxwell} spacetimes},
  Author                   = {Paul D. Lasky and Anthony W. C. Lun},
  Journal                  = prd,
  Year                     = {2007},
  Pages                    = {104010},
  Volume                   = {75},

  Archiveprefix            = {arXiv},
  Doi                      = {10.1103/PhysRevD.75.104010},
  Eprint                   = {0704.3634},
  Primaryclass             = {gr-qc}
}

@Article{LaskyLun06b,
  Title                    = {Generalized {Lemaitre-Tolman-Bondi} solutions with pressure},
  Author                   = {P. D. Lasky and A. W. C. Lun},
  Journal                  = prd,
  Year                     = {2006},

  Month                    = oct,
  Pages                    = {084013},
  Volume                   = {74},

  Archiveprefix            = {arXiv},
  Doi                      = {10.1103/PhysRevD.74.084013},
  Eprint                   = {gr-qc/0606055},
  Issue                    = {8},
  Numpages                 = {10},
  Publisher                = {American Physical Society}
}

@Article{LeDMM09a,
  Title                    = {Role of shell crossing on the existence and stability of trapped matter shells in spherical inhomogeneous $\Lambda$-CDM models},
  Author                   = {{Le Delliou}, Morgan and Filipe C. Mena and Jos\'{e} P. Mimoso},
  Journal                  = prd,
  Year                     = {2011},

  Month                    = may,
  Pages                    = {103528},
  Volume                   = {83},

  Archiveprefix            = {arXiv},
  Doi                      = {10.1103/PhysRevD.83.103528},
  Eprint                   = {1103.0976},
  Issue                    = {10},
  Numpages                 = {16},
  Primaryclass             = {gr-qc},
  Publisher                = {American Physical Society}
}

@Article{Lemaitre:1933gd,
  Title                    = {The expanding universe},
  Author                   = {G. Lema\^{\i}tre},
  Journal                  = aasba,
  Year                     = {1933},
  Note                     = {Reprinted in \href{http://dx.doi.org/10.1023/A:1018855621348}{\grg{} \textbf{29}(5), 641--680 (1997)}},
  Pages                    = {51--85},
  Volume                   = {53}
}

@Article{Mimoso:2013iga,
  Title                    = {Local conditions separating expansion from collapse in spherically symmetric models with anisotropic pressures},
  Author                   = {Jos\'{e} P. Mimoso and {Le Delliou}, Morgan and Filipe C. Mena},
  Journal                  = prd,
  Year                     = {2013},

  Month                    = aug,
  Pages                    = {043501},
  Volume                   = {88},

  Archiveprefix            = {arXiv},
  Doi                      = {10.1103/PhysRevD.88.043501},
  Eprint                   = {1302.6186},
  Issue                    = {4},
  Numpages                 = {16},
  Primaryclass             = {gr-qc},
  Publisher                = {American Physical Society}
}

@Article{MLeDM09,
  Title                    = {Separating expansion from contraction in spherically symmetric models with a perfect fluid: Generalization of the Tolman-Oppenheimer-Volkoff condition and application to models with a cosmological constant},
  Author                   = {Jos\'e P. Mimoso and {Le Delliou}, Morgan and Filipe C. Mena},
  Journal                  = prd,
  Year                     = {2010},

  Month                    = jun,
  Pages                    = {123514},
  Volume                   = {81},

  Archiveprefix            = {arXiv},
  Doi                      = {10.1103/PhysRevD.81.123514},
  Eprint                   = {0910.5755},
  Issue                    = {12},
  Numpages                 = {17},
  Primaryclass             = {gr-qc},
  Publisher                = {American Physical Society}
}

@Article{MisnerSharp,
  Title                    = {Relativistic Equations for Adiabatic, Spherically Symmetric Gravitational Collapse},
  Author                   = {Charles W. Misner and David H. Sharp},
  Journal                  = physrev,
  Year                     = {1964},

  Month                    = oct,
  Pages                    = {B571--B576},
  Volume                   = {136},

  Doi                      = {10.1103/PhysRev.136.B571},
  Issue                    = {2B},
  Publisher                = {American Physical Society}
}

@Article{navarro-1996,
  Title                    = {The Structure of Cold Dark Matter Halos},
  Author                   = {J. F. Navarro and C. S. Frenk and S. D. M. White},
  Journal                  = apj,
  Year                     = {1996},

  Month                    = may,
  Pages                    = {563--575},
  Volume                   = {462},

  Doi                      = {10.1086/177173},
  Eprint                   = {astro-ph/9508025},
  Primaryclass             = {arXiv}
}

@Article{oppenheimer-1939a,
  Title                    = {On Massive Neutron Cores},
  Author                   = {J. R. Oppenheimer and G. M. Volkoff},
  Journal                  = physrev,
  Year                     = {1939},

  Month                    = feb,
  Pages                    = {374--381},
  Volume                   = {55},

  Doi                      = {10.1103/PhysRev.55.374},
  Issue                    = {4},
  Publisher                = {American Physical Society}
}

@Article{Tolman:1934za,
  Title                    = {Effect of Inhomogeneity on Cosmological Models},
  Author                   = {Richard C. Tolman},
  Journal                  = pnas,
  Year                     = {1934},

  Month                    = mar,
  Pages                    = {169--176},
  Volume                   = {20},
  Adsurl                   = {http://adsabs.harvard.edu/abs/1934PNAS...20..169T},
  Doi                      = {10.1073/pnas.20.3.169}
}

@article{Gunn:1972sv,
    author = "Gunn, James E. and Gott, III, J. Richard",
    title = "{On the Infall of Matter into Clusters of Galaxies and Some Effects on Their Evolution}",
    doi = "10.1086/151605",
    journal = "Astrophys. J.",
    volume = "176",
    pages = "1--19",
    year = "1972"
}

@article{Bertschinger:1985pd,
    author = "Bertschinger, E.",
    title = "{Self - similar secondary infall and accretion in an Einstein-de Sitter universe}",
    doi = "10.1086/191028",
    journal = "Astrophys. J. Suppl.",
    volume = "58",
    pages = "39",
    year = "1985"
}

@article{Fillmore:1984wk,
    author = "Fillmore, J. A. and Goldreich, P.",
    title = "{Self-similiar gravitational collapse in an expanding universe}",
    doi = "10.1086/162070",
    journal = "Astrophys. J.",
    volume = "281",
    pages = "1--8",
    year = "1984"
}

@article{DelPopolo:2020mge,
    author = "Del Popolo, Antonino and Chan, Man Ho",
    title = "{Turnaround radius in $\Lambda$CDM and dark matter cosmologies. II. The role of dynamical friction}",
    eprint = "2104.12768",
    archivePrefix = "arXiv",
    primaryClass = "astro-ph.CO",
    doi = "10.1103/PhysRevD.102.123510",
    journal = "Phys. Rev. D",
    volume = "102",
    number = "12",
    pages = "123510",
    year = "2020"
}

@article{Korkidis:2019nzk,
    author = "Korkidis, Giorgos and Pavlidou, Vasiliki and Tassis, Konstantinos and Ntormousi, Evangelia and Tomaras, Theodore N. and Kovlakas, Konstantinos",
    title = "{Turnaround radius of galaxy clusters in N-body simulations}",
    eprint = "1912.08216",
}

@ARTICLE{2019arXiv191203687L,
       author = {{Lahav}, Ofer and {Liddle}, Andrew R},
        title = "{The Cosmological Parameters (2019)}",
      journal = {arXiv e-prints},
         year = 2019,
        month = dec,
          eid = {arXiv:1912.03687},
        pages = {arXiv:1912.03687},
archivePrefix = {arXiv},
       eprint = {1912.03687},
 primaryClass = {astro-ph.CO},
       adsurl = {https://ui.adsabs.harvard.edu/abs/2019arXiv191203687L}
}

@article{Gautreau:1984PhRvD186,
  title = {Curvature coordinates in cosmology},
  author = {Gautreau, Ronald},
  journal = {Phys. Rev. D},
  volume = {29},
  issue = {2},
  pages = {186--197},
  numpages = {0},
  year = {1984},
  month = {Jan},
  publisher = {American Physical Society},
  doi = {10.1103/PhysRevD.29.186},
  url = {https://link.aps.org/doi/10.1103/PhysRevD.29.186}
}

@article{DiPrisco:1997tw,
    author = "Di Prisco, A. and Herrera, L. and Varela, V.",
    title = "{Cracking of Homogeneous Self-Gravitating Compact Objects Induced by Fluctuations of Local Anisotropy}",
    doi = "10.1023/A:1018859712881",
    journal = "Gen. Rel. Grav.",
    volume = "29",
    pages = "1239--1256",
    year = "1997"
}

@article{Ringstrom:2015jza,
    author = {Ringstr{\"o}m, Hans},
    title = "{Origins and development of the Cauchy problem in general relativity}",
    doi = "10.1088/0264-9381/32/12/124003",
    journal = "Class. Quant. Grav.",
    volume = "32",
    number = "12",
    pages = "124003",
    year = "2015"
}

@inproceedings{Bartnik:2002cw,
    author = "Bartnik, Robert and Isenberg, Jim",
    title = "{The Constraint equations}",
    booktitle = "{50 Years of the Cauchy Problem in General Relativity: Summer School on Mathematical Relativity and Global Properties of Solutions of Einstein's Equations}",
    eprint = "gr-qc/0405092",
    archivePrefix = "arXiv",
    doi = "10.1007/978-3-0348-7953-8_1",
    year = "2002"
}

@article{Chrusciel:2004cc,
    author = "Chrusciel, Piotr T. and Isenberg, James and Pollack, Daniel",
    title = "{Initial data engineering}",
    eprint = "gr-qc/0403066",
    archivePrefix = "arXiv",
    doi = "10.1007/s00220-005-1345-2",
    journal = "Commun. Math. Phys.",
    volume = "257",
    pages = "29--42",
    year = "2005"
}

@article{Anderson:2000mt,
    author = "Anderson, Michael T.",
    title = "{On long time evolution in general relativity and geometrization of three manifolds}",
    eprint = "gr-qc/0006042",
    archivePrefix = "arXiv",
    doi = "10.1007/s002200100527",
    journal = "Commun. Math. Phys.",
    volume = "222",
    pages = "533--567",
    year = "2001"
}

@article{Choquet-Bruhat:1969ywq,
    author = "Choquet-Bruhat, Y. and Geroch, Robert P.",
    title = "{Global aspects of the Cauchy problem in general relativity}",
    doi = "10.1007/BF01645389",
    journal = "Commun. Math. Phys.",
    volume = "14",
    pages = "329--335",
    year = "1969"
}

@article{Foures-Bruhat:1952grw,
    author = "Foures-Bruhat, Y.",
    title = "{Theoreme d'existence pour certains systemes derivees partielles non lineaires}",
    doi = "10.1007/BF02392131",
    journal = "Acta Mat.",
    volume = "88",
    pages = "141--225",
    year = "1952"
}

@article{Komar:1958ymq,
    author = "Komar, Arthur",
    title = "{Construction of a Complete Set of Independent Observables in the General Theory of Relativity}",
    doi = "10.1103/PhysRev.111.1182",
    journal = "Phys. Rev.",
    volume = "111",
    number = "4",
    pages = "1182",
    year = "1958"
}

@article{Deser:1967zzb,
    author = "Deser, Stanley",
    title = "{Covariant decomposition of symmetric tensors and the gravitational Cauchy problem}",
    journal = "Ann. Inst. H. Poincare Phys. Theor. A",
    volume = "7",
    number = "2",
    pages = "149--188",
    year = "1967"
}

@inproceedings{York:2004gb,
    author = "York, Jr., James W.",
    title = "{The Initial value problem using metric and extrinsic curvature}",
    booktitle = "{10th Marcel Grossmann Meeting on Recent Developments in Theoretical and Experimental General Relativity, Gravitation and Relativistic Field Theories (MG X MMIII)}",
    eprint = "gr-qc/0405005",
    archivePrefix = "arXiv",
    doi = "10.1142/9789812704030_0001",
    pages = "3--16",
    month = "5",
    year = "2004"
}

@article{York:1972sj,
    author = "York, Jr., James W.",
    title = "{Role of conformal three geometry in the dynamics of gravitation}",
    doi = "10.1103/PhysRevLett.28.1082",
    journal = "Phys. Rev. Lett.",
    volume = "28",
    pages = "1082--1085",
    year = "1972"
}

@article{York:1973ia,
    author = "York, Jr., James W.",
    title = "{Conformatlly invariant orthogonal decomposition of symmetric tensors on Riemannian manifolds and the initial value problem of general relativity}",
    doi = "10.1063/1.1666338",
    journal = "J. Math. Phys.",
    volume = "14",
    pages = "456--464",
    year = "1973"
}

@article{Lichnerowicz:1944zz,
    author = "Lichnerowicz, A.",
    title = "{On completely harmonic Riemannian spaces}",
    journal = "Bull. Soc. Math. Fr.",
    volume = "72",
    pages = "146--168",
    year = "1944"
}

@article{Lichnerowicz1944,
  author  = {Lichnerowicz, Andr{\'e}},
  title   = {L'int{\'e}gration des {\'e}quations de la gravitation relativiste et le probl{\`e}me des $n$ corps},
  journal = {Journal de Math{\'e}matiques Pures et Appliqu{\'e}es},
  series  = {9e s{\'e}rie},
  volume  = {23},
  pages   = {37--63},
  year    = {1944},
  language = {french},
  url     = {http://www.numdam.org/item/JMPA_1944_9_23__37_0/}
}

@article{York:1971hw,
    author = "York, Jr., James W.",
    title = "{Gravitational degrees of freedom and the initial-value problem}",
    doi = "10.1103/PhysRevLett.26.1656",
    journal = "Phys. Rev. Lett.",
    volume = "26",
    pages = "1656--1658",
    year = "1971"
}

@inbook{Isenberg:2013iva,
    author = "Isenberg, James",
    editor = "Ashtekar, Abhay and Petkov, Vesselin",
    title = "{The Initial Value Problem in General Relativity}",
    booktitle = "{Springer Handbook of Spacetime}",
    eprint = "1304.1960",
    archivePrefix = "arXiv",
    primaryClass = "gr-qc",
    doi = "10.1007/978-3-642-41992-8_16",
    pages = "303--321",
    year = "2014"
}

@article{Tichy:2016vmv,
    author = "Tichy, Wolfgang",
    title = "{The initial value problem as it relates to numerical relativity}",
    eprint = "1610.03805",
    archivePrefix = "arXiv",
    primaryClass = "gr-qc",
    doi = "10.1088/1361-6633/80/2/026901",
    journal = "Rept. Prog. Phys.",
    volume = "80",
    number = "2",
    pages = "026901",
    year = "2017"
}

@article{Gourgoulhon:2007ue,
    author = "Gourgoulhon, Eric",
    title = "{3+1 formalism and bases of numerical relativity}",
    eprint = "gr-qc/0703035",
    archivePrefix = "arXiv",
    month = "3",
    year = "2007"
}

@preamble{"\newcommand{\grg}{Gen.\ Relativ.\ Gravit.\@}"}

@string{aasba = "Annales Soc.\ Sci.\ Brux.\ Ser.\ A"}

@string{apj = "Astrophys.\ J.\@"}

@string{grg = "\grg"}

@string{mn = "Mon.\ Not.\ R.\ Astron.\ Soc.\@"}

@string{physrev = "Phys.\ Rev.\@"}

@string{prd = "Phys.\ Rev.\ D"}

@string{pla = "Phys.\ Lett.\ A"}

@string{pnas = "Proc.\ Natl.\ Acad.\ Sci. U.S.A.\@"}
\end{document}